\documentclass[a4paper,12pt]{article}
\usepackage{epsfig}
\usepackage{amssymb}
\usepackage{amsfonts}
\usepackage{amsmath}
\usepackage{color}

\numberwithin{equation}{section}

\newcommand{\bea}{\begin{eqnarray}}
\newcommand{\eea}{\end{eqnarray}}

\def \d{\textrm{d}}
\def\Tr{{\rm Tr}\,}
\def\F{\mathcal{F}}
\def\A{\mathcal{A}}
\def\eps{\epsilon}

\def\DaD{${\rm D7-}{\overline{\rm D7}}$\ }
\def\DaDe{${\rm D8-}{\overline{\rm D8}}$\ }

\newcommand{\Z}{\ensuremath{\mathbb Z}}

\begin{document}

\thispagestyle{empty}

\begin{flushright}
ITEP-TH-45/10 \\
TAUP-2925-10
\end{flushright}

\vspace{18pt}
\begin{center}
{\Large \textbf{Attractive Holographic Baryons}}
\end{center}

\vspace{6pt}
\begin{center}
{\large{Anatoly Dymarsky$^{a,b}$, Dmitry Melnikov$^{b,c}$ and Jacob Sonnenschein$^c$ }\\}
\vspace{25pt}
\textit{\small $^a$ School of Natural Sciences, Institute for Advanced Study,\\Princeton, NJ 08540, USA}\\ \vspace{6pt}
\textit{\small $^b$  Institute for Theoretical and Experimental Physics, \\B.~Cheremushkinskaya 25, Moscow 117218, Russia}\\ \vspace{6pt}
\textit{\small $^c$ The Raymond and Beverly Sackler School of Physics and Astronomy, \\ Tel Aviv University, Ramat Aviv 69978, Israel}\\ \vspace{6pt}
\end{center}

\vspace{12pt}

\begin{abstract}

We propose a holographic model of baryon interactions  based on non-supersymmetric \DaD flavor branes embedded in the Klebanov-Strassler background. The baryons are D3-branes wrapping the $S^3$ of the conifold  with  $M$ strings connecting the D3 and the flavor branes.  Depending on the location of the latter there are two possibilities: the D3 either remains separate from the flavor branes or dissolves in them and becomes a flavor instanton. The leading order interaction between the baryons is a competition between the attraction and the repulsion due to the $\sigma$ and $\omega$ mesons. The lightest  $0^{++}$ particle $\sigma$ is a pseudo-Goldstone boson associated with the spontaneous breaking of scale invariance. In a certain range of parameters it is parametrically lighter than any other massive state. As a result at large distances baryons attract each other. At short distances the potential admits a repulsive core due to an exchange of the  $\omega$ vector meson. We discuss baryon coupling to glueballs, massive mesons and pions and point out the  condition for the model to have a small binding energy.

\end{abstract}

\vspace{4pt} {\small

\noindent }

\newpage

\section{Introduction and overview}

Understanding nuclear forces is an important problem with many interesting questions still not answered. One of them is the smallness of the nuclear binding energy. Despite the strength of the strong interactions the force between two nucleons is rather weak: the binding energy is only a few percent of the typical baryon mass. At long distance the force is mediated by exchanges of the lightest isoscalar particles:\footnote{We implicitly assume the case of baryonic matter when the contribution due to one-pion exchange is believed to average to zero.} $\sigma$ ($0^{++}$) and $\omega$ ($1^{--}$). The masses of $\sigma$ and $\omega$ are close, hence they effectively cancel each other's contribution to the binding energy. The origin of such a coincidence is not easy to establish. Moreover, the identification of the scalar meson and its relation to a pion-pion resonance is a long-standing puzzle. Theoretically one may also worry about the glueballs though they are heavier than  $\sigma$ and in many cases can be neglected. Accounting for these phenomena using first principles of QCD has been found to be a non-tractable task.

Provided the approximate equality of $\sigma$ and $\omega$ masses is not totally coincidental, to find a reason it would help to study a toy model, in which the contributing factors can be considered separately. A natural idea would be to look for a holographic model of interacting baryons, in particular because in the large $N_c$ limit, required by the holographic approach, the exchange by many particles as well as by glueballs will be suppressed. This would leave to understand whether the contribution from the single scalar and vector meson exchanges approximately cancel each other. Attempts have been made recently to find such a holographic model. In~\cite{KimZahed,Hashimoto,Kaplunovsky} the baryons were considered in the context of the Sakai-Sugimoto model~\cite{SakaiSugimoto}. It turns out that in this model the attractive force between the baryons is always weaker than the repulsive one~\cite{Kaplunovsky}. The light scalar meson happens to be heavier than the lightest vector meson resulting in a net repulsive force. Clearly this is very different from hadronic physics. In order to shed some light   on  the  nucleon  binding energy and in particular its smallness,  we should first try to find a model featuring a binding potential and then try to examine whether it is naturally small.

Motivated by the above considerations, in this paper we study a holographic model based on \DaD branes of \cite{KSII} embedded in the Klebanov-Strassler (KS) background~\cite{KS}. This system is scale invariant in the UV. Adding the \DaD branes spontaneously break conformal symmetry by a vev of a marginal operator at some scale $r_0$. When this scale is larger than the internal scale of the gauge theory $r_\epsilon\equiv \epsilon^{2/3}$, the lightest scalar meson is parametrically light as a pseudo-Goldstone boson of the conformal symmetry. This meson gives the leading contribution to the attractive force and we will retain the notation $\sigma$ for it in the remainder of the paper. Similarly, we will call $\omega$ the lightest vector meson, which corresponds to the fluctuations of the $U(1)$ gauge field on the world-volume of the \DaD branes.

The model in question has the following hierarchy of light particles. We are working in the quenched approximation, that is the number of \DaD branes $N_f$ is much smaller than the number of colors $M$. Hence we ignore the backreaction of the open string sector. In particular, the mass of glueballs remains the same as in the KS theory and therefore is $r_0$-independent. The typical scale of the glueball mass is
\bea
m_{\text{gb}} \sim \frac{r_\epsilon}{\alpha'\lambda}\, , \quad \lambda=g_s M\, .
\eea
In the regime $r_0\gg r_\epsilon$ the theory is (almost) conformal and therefore the mass of mesons can depend
only on the scale of symmetry breaking $r_0$
\bea
m_{\rm meson}\sim \frac{r_0}{\alpha'\lambda}= {r_0\over r_\epsilon }\,m_{\rm gb}\, .
\eea
The pseudo-Goldstone boson $\sigma$ is parametrically lighter
\bea
m_{\sigma}\sim  {r^2_\epsilon \over r_0^2}\,m_{\rm gb}\, .
\eea
We will explain this result later in the text.

As $r_0$ approaches $r_\epsilon$ the mesons become lighter, while the pseudo-Gold\-stone grows heavier. Around the minimal value $r_0=r_\epsilon$ all mesons have approximately the same mass of order $m_{\rm gb}$. This is the most interesting regime of parameters because for $r_0\sim r_\epsilon$ the approximate cancelation of the attractive and the repulsive force can occur naturally. Hence it is crucial to know the precise hierarchy of the lightest particles in this limit. Such a detailed computation is beyond the scope of the present work. A recent publication~\cite{Ihl} based on the approximate analysis ignoring an intricate dynamics of the gauge fields on \DaD branes claims the following spectrum in the limit $r_0=r_\epsilon$
\bea
m_{\sigma} < m_{0^{++}} < m_{\omega} < m_{1^{++}}\,.
\eea
Here $0^{++}$ and $1^{++}$ denote the lightest glueballs with the corresponding quantum numbers and we have used the results on the KS glueballs from~\cite{Berg,Gordeli}. This result is encouraging and we will investigate the phenomenology of the model assuming that $m_\sigma<m_{\omega}$ for $r_0\sim r_\epsilon$.

A baryon in our setup is  represented by  a D3-brane wrapping the $S^3$ of the conifold and a set of $M$ strings  connecting it to the \DaD branes. For large values of the parameter $r_0\gg r_\epsilon$ the string tension is smaller than the force exerted on D3 due to curved geometry. To minimize  the energy D3-brane will settle near the tip of the conifold at $r\sim r_\epsilon$ with the D3$-$D7 strings stretched all the way between $r$ and $r_0$. When $r_0$ is significantly close to $r_\epsilon$ the geometry can be effectively approximated by the flat one and creates only a mild force. The string tension wins, and the D3-brane is pulled towards the \DaD branes and dissolves there becoming an instanton.

The description of the baryons in the two regimes is essentially different. Nevertheless, in both regimes the D3-brane is close to the tip of the conifold and the form of the glueball induced potential is approximately the same
\bea
\label{Vgb intro}
V_{\rm gb}\sim \pm {\text{e}^{-m_{\rm gb}|x|}\over |x|}\, .
\eea
Here $x$ is the 3-dimensional distance between  the two baryons which is assumed to be much larger than the size of the baryons $\rho$. In the above calculation we estimated the force at large distances, where it is dominated by the lightest particles. In this approximation baryon is treated as a localized source. The sign in~(\ref{Vgb intro}) depends on the spin of the exchanged particle: the scalar induces attraction while the vector yields repulsion. In the KS theory the lightest $0^{++}$ glueball is lighter than the vector $1^{++}$ and the coupling constant can in principle be read off the explicit form of the wavefunction~\cite{Berg,Gordeli}. Although the glueballs will generate some potential, this fact will not be important for us, since the glueball potential is suppressed by $g_s$ as compared to the potential generated by mesons.

Let us first focus on the regime $r_0\gg r_\epsilon$, in which one adopts the string picture of the baryon. The main interaction at large distances is due to the lightest mode $\sigma$. The effective potential of the scalar meson exchange reads
\bea
\label{Vs intro}
V_{\rm scalar}\sim -{1\over g_s \log (r_0/r_\epsilon)}{\text{e}^{-m_{\sigma}|x|}\over |x|}\, .
\eea
It is indeed parametrically larger than~(\ref{Vgb intro}) in the holographic regime $g_s\to 0$. More precisely, to ignore glueballs we need to satisfy $g_s\log (r_0/r_\epsilon)\ll 1$.

Since the $\omega$-meson is much heavier than $\sigma$ in the $r\gg r_\eps$ regime, the repulsive interaction is subdominant at large distances, although the potential is similar to~(\ref{Vs intro})
\bea
\label{Vv intro}
V_{\rm vector}\sim  {1\over g_s \log (r_0/r_\epsilon)}{\text{e}^{-m_{\omega}|x|}\over |x|}\, .
\eea

Theoretically it is possible that in this regime the potential will stay attractive even at distances $|x|\ll m_{\omega}^{-1}$. Phenomenologically this is not an appealing scenario.  Assuming  that the repulsion will start winning at some $|x|\sim m_{\omega}^{-1}$ this will imply the binding energy of the order
\bea
\label{Ebind large}
E_{\rm binding}\sim {1\over g_s}\, m_{\rm vector} \sim {r_0\over g_s^2 M \alpha'}\, .
\eea
This is parametrically smaller than the baryon mass,
\bea
m_{\rm baryon}={r_\epsilon M\over \alpha'}\, ,
\eea
but the relative smallness is due to the artifact of the holographic approach -- the large 't Hooft constant $\lambda =g_s M \gg 1$, not a cancelation between the attraction and repulsion.

Now let us consider the opposite regime $r_0\rightarrow r_\epsilon$. For $r_0$ small enough  the wrapped D3-brane will dissolve in the \DaD and will be described by an instanton. We will assume that the scalar $\sigma$ is still lighter than the vector $\omega$ in this regime. When $r_0$ reaches its minimal value $r_\epsilon$, the resulting configuration of the \DaD branes is invariant under an emergent $U(1)$ symmetry. The wavefunction of $\sigma$ is odd under $\mathbb{Z}_2\in U(1)$ and therefore the leading coupling of $\sigma$ to baryons vanishes. The same phenomena also occurs in the Sakai-Sugimoto model~\cite{Kaplunovsky}. By varying $r_0$ near the point $r_0=r_\epsilon$  one can tune the coupling of $\sigma$ to be small.

The net potential in this case can be written in the form (cf.~\cite{Walecka})
\bea
V = \frac{1}{4\pi g_s}\left( g_\omega^2\,{\text{e}^{-m_{\omega} |x|}\over |x|} - g_\sigma^2\, {\text{e}^{-m_{\sigma} |x|}\over |x|} \right).
\eea
It is valid only for $|x|$ large enough. Here $g_\sigma$ and $g_\omega$ encode the coupling of the mesons to the baryons and we have kept the factor of $g_s$ explicit. If $m_\sigma< m_\omega$, the potential is attractive at large distances no matter what the couplings are. On the other hand if $g_\sigma$ is small enough, at distances shorter than $m_{\omega}^{-1}$ the vector interaction ``wins'' and the potential becomes repulsive. The binding energy
\bea
E_{\rm binding}\sim  \kappa {r_\epsilon \over g_s^2 M \alpha'}\, ,
\eea
is suppressed by a small dimensionless number $\kappa$, which is related to the smallness of the coupling $g_\sigma$ and the fact that $m_{\sigma}$ and $m_{\omega}$ are of the same order. This suppression is in addition to the relative $\lambda^{-2}$ factor (compared with the baryon mass). The extra factor $\kappa$ is phenomenologically promising as it represents the near-cancelation of the attractive and repulsive forces responsible for the small binding energy in hadron physics.

This paper is organized as follows. In section~\ref{massesofmesons} we describe the setup and determine  the mass scales of the  mesons and glueballs. In section~\ref{svi} we introduce a baryon into the D3$-$D7 geometry and discuss its equilibrium positions. There are two limiting cases of interest: for large $r_0$ the baryon sits at the tip of the conifold, while for $r\sim \epsilon$ it dissolves into \DaD and becomes an instanton. In section~\ref{glueballs} we investigate the relevance of the glueballs to the interaction of the two baryons. We find that it is suppressed by $g_s$ compared with the interaction carried by the mesons, which we study in section~\ref{mesons}. We consider two regimes, corresponding to a stringy and  an instanton pictures of baryons. Our analysis is summarized in section~\ref{conclusions}, where we also discuss the implications of the model and open questions. In the appendices we review the question of the stability of the \DaD embedding~(appendix \ref{apndx stability}) and obtain estimates for the mass and the size of the baryon~(appendix \ref{sizeofbaryon}).


\section{Mesons}
\label{massesofmesons}

\subsection{The model}
Let us start with the description of  the holographic setup. To simplify the presentation we will frequently compare our model with the more familiar one of Sakai and Sugimoto~\cite{SakaiSugimoto}. The KS geometry results from $N$ D3-branes and $M$ fractional D3-branes placed on the conical singularity of a conifold, a cone over the $T^{1,1}$ space.\footnote{ $T^{1,1}\simeq S^3\times S^2$ and fractional D3-branes are the D5-branes wrapped on $S^2$ of $T^{1,1}$.} The dual gauge theory is a ${\cal{N}}=1$ SYM with the gauge group $SU(N+M)\times SU(N)$ and bifundamental matter in a left and right doublets of the global $SU(2)\times SU(2)$ symmetry. The theory flows to the IR by undergoing a chain of Seiberg dualities and ends as a confining $SU(M)$ SYM. The light states in the low-energy theory are super-glueballs and hybrids containing bifundamentals. They were extensively studied in the $SU(2)\times SU(2)$ singlet sector~\cite{Berg,Gordeli,Krasnitz,BDKMS} and beyond~\cite{Bianchi,Pufu}.

The Sakai-Sugimoto model is based on the $N_f$ D8 and $\overline{\text{D8}}$ flavor branes immersed in the geometry of  $N_c$ D4-branes compactified on a circle with anti-periodic boundary conditions~\cite{WittensModel}. The D8 and $\overline{\text{D8}}$-branes smoothly connect to form U-shape configurations thus realizing the dynamical chiral symmetry breaking $U(N_f)\times U(N_f)\rightarrow U(N_f)$. Similar U-shape configurations of \DaD branes in the context of the conifold were first proposed in~\cite{KSII} for the Klebanov-Witten (KW) geometry~\cite{KW}, a conformal $M\to 0$ limit of the KS model. They were extended to the KS case in~\cite{DKSII}. Now we will review this construction in more detail.

The original  and fractional D3-branes are placed at the tip of the conifold spanning the Minkowski directions. By analogy with the Sakai-Sugimoto model four of the eight directions of the D7 and $\overline{\text{D7}}$ are in Minkowski, three are wrapping the $S^3$ of the $T^{1,1}$ and the remaining one is a line in  $R^+\times S^2$. The geometry of the profile is controlled by the effective DBI action
\bea
\label{DBI}
\displaystyle
S = - \frac{1}{(2\pi)^7{\alpha'}^{4}} \int_{\text{D7}+\overline{\text{D7}}} \d^{8}x\,\text{e}^{-\phi}\sqrt{-\det(g_8 + 2\pi\alpha'\F)}\,,
\eea
where $g_8$ is the induced metric and $\F$ is the $U(N_f)$ gauge field on the world-volume of the D7 and $\overline{\text{D7}}$. The dilaton is constant $\text{e}^{\phi}=g_s$, unlike in the  Sakai-Sugimoto model.

\begin{figure}[htb]
\begin{minipage}[b]{0.5\linewidth}
\begin{center}

\includegraphics[width=6.7cm]{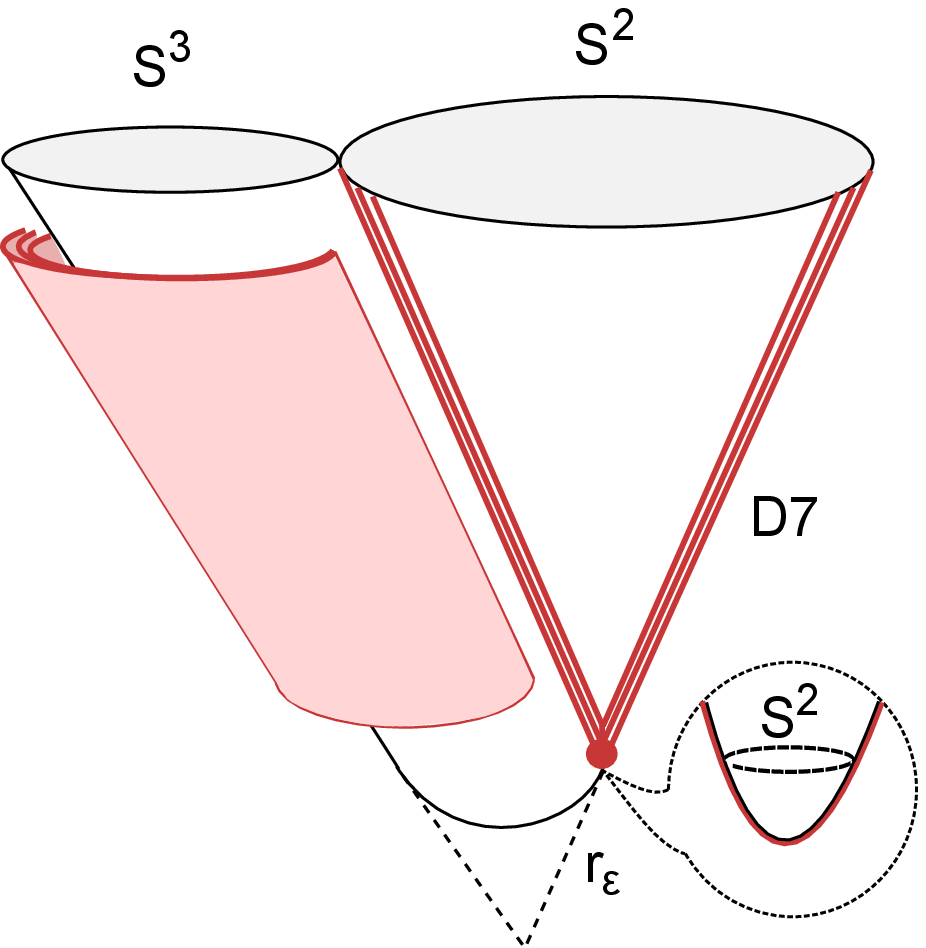}

(a)
\end{center}
\end{minipage}
\begin{minipage}[b]{0.5\linewidth}
\begin{center}

\includegraphics[width=6.5cm]{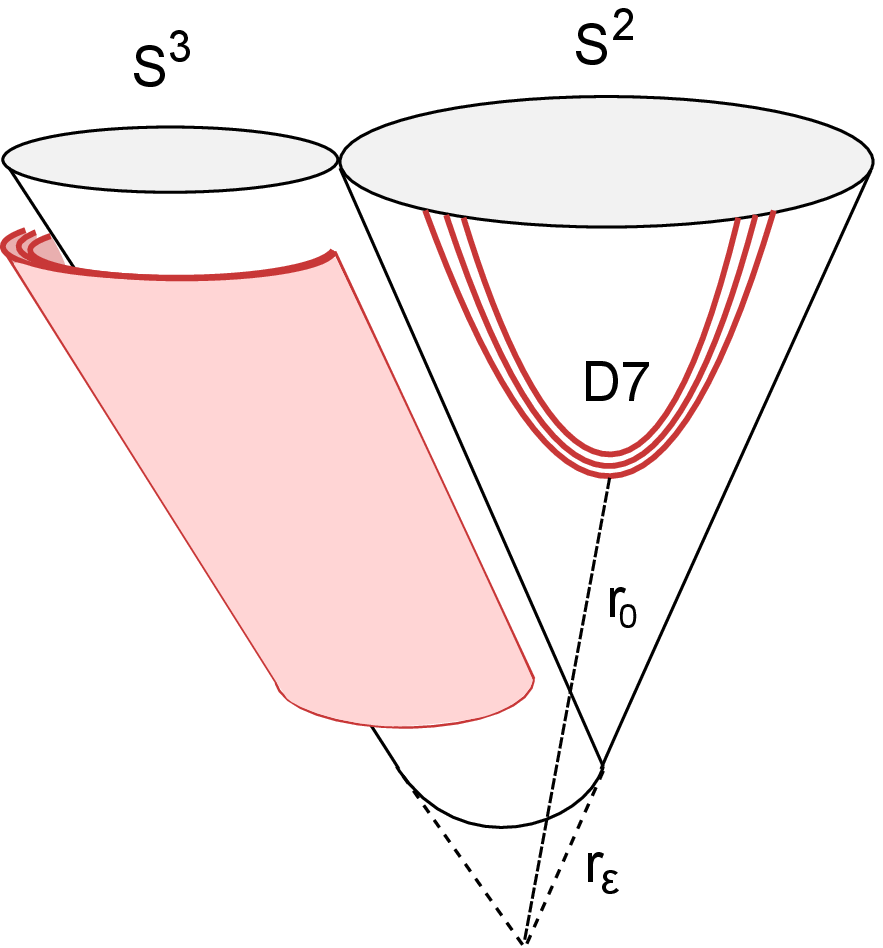}

(b)
\end{center}
\end{minipage}
\vspace{-0.6cm}
\caption{\small Embedding of the D7-branes. The branes wrap $S^3$ and span a line in $R^+\times S^2$. The lowest point of the \DaD profile is at $r_0$. In the antipodal case (a) the configuration descends all the way down to the tip of the conifold $r_0=r_\eps$. It is smooth at the tip. In the non-antipodal case (b) the lowest point of \DaD is at some distance from the tip,  $r_0>r_\eps$.}
\label{fig embedding}
\end{figure}

In both the KW case~\cite{KSII} and the KS case~\cite{DKSII} the position of the D7 and $\overline{\text{D7}}$ on the $R^+\times S^2$ can be restricted to the equator $S^1$ of $S^2$ by an appropriate $SU(2)$ symmetry transformation. Effectively one obtains a U-shaped profile of the connected D7 and $\overline{\text{D7}}$ pairs on a cigar-like manifold $R^+\times S^1$, see figure~{\ref{fig embedding}}.

To make a better connection with the Sakai-Sugimoto construction we map the parameter of the two models as follows. The $M_{KK}$ mass scale of the Sakai-Sugimoto model, which is related to the coordinate distance $U_{KK}$ to the cigar from the origin, is mapped to the KS scale $r_\epsilon\equiv\epsilon^{2/3}$. The latter is, roughly speaking, a similar distance measuring the deviation of the deformed conifold from the singular one. The minimal distance $U_0$ between the \DaDe branes and the and the origin $U=0$ is mapped to the minimal distance from the origin to the \DaD branes $r_0$. The case $r_0=r_\epsilon$ is the antipodal configuration, for which the position on the $S^{1}$ of the cigar (equator of the $S^2$) is the same for all $r$.

In the KW case the regularized energy of the \DaD configuration does not depend on $r_0$. Therefore there is a massless particle corresponding to variations of $r_0$. It is a Goldstone boson of the spontaneously broken scaling symmetry. The symmetry is broken dynamically in the IR, rather than by the boundary conditions in the UV. Indeed the asymptotic angular separation along the equator $\Delta\phi(r_\infty)=\sqrt{6}\pi/4$ is the same for all $r_0$. The angle $\phi$ is dimensionless and the boundary conditions at infinity $\phi\to {\rm const}$ do not break conformal symmetry. This is in contrast with the Kuperstein embedding~\cite{Kuperstein} (the KS analog of the Karch-Katz embedding~\cite{KK}) specified by the dimensionful parameter $z_4=\mu$.

In the KS case the conformal symmetry is also broken explicitly by the geometry. Therefore the variation of $r_0$ is no longer massless and becomes a pseudo-Goldstone boson. As we demonstrate below, the pseudo-Goldstone boson can be made significantly lighter than other mesons and glueballs.

Before proceeding to the discussion on the particle spectra in the next section we will briefly discuss the unbroken symmetries of the model. The original KS geometry has an $SU(2)\times SU(2)$ symmetry that rotates $T^{1,1}$. The left $SU(2)$ is preserved by \DaD as it rotates the $S^3$ of the conifold, which the \DaD branes wrap. The right $SU(2)$ rotates the $S^2$ and is therefore broken. In the antipodal case the $U(1)$ of the second $SU(2)$ survives with important consequences for the phenomenology of the model.

Another symmetry of the KS solution is the ${\cal{I}}$-symmetry: a $\mathbb{Z}_2$ which interchanges the two $S^2$ of the conifold with simultaneous flipping of the $F_3$ and $H_3$ signs~\cite{KW}. In the dual field theory this symmetry acts as charge conjugation combined with the interchanging  of the bifundamentals. The \DaD branes will break this symmetry, because they wrap a particular $S^3$. For a generic $r$, the $S^3$ is not symmetric with respect to the swap of the two-spheres, except for the point at the tip of the conifold.

The \DaD branes break supersymmetry and may significantly change the KS model through a backreaction. In this paper we assume the quenched approximation $N_f\ll M$ when the number of flavor branes is small. Thus we can ignore the backreaction of the \DaD\!\!, in which case the glueball sector remains approximately supersymmetric, while the flavor sector has broken supersymmetry. To simplify the presentation we sometimes will omit the $N_f$ factor in what follows, since the dependence on $N_f$ can be easily restored.


\subsection{Masses of vector and scalar mesons}
In this section we will estimate the masses of lightest vector and scalar mesons in the models of~\cite{KSII,DKSII} and show that the pseudo-Goldstone boson $\sigma$ can be made parametrically lighter than the vector $\omega$. This is in a stark contrast with the Sakai-Sugimoto model, in which $\sigma$ is always heavier than $\omega$~\cite{Mintkevich}. To see how this works we start with the conformal background of the KW solution and embed the \DaD branes as described in the previous section. The embedding can be parameterized by $\phi(r)$, namely, the  functional dependence  of the angle along the equator of $S^2$ on the radius. The effective action for $\phi(r)$ has the form~\cite{KSII}
\bea
\label{effS}
S={1\over g_s \alpha'^4}\int r^3 \sqrt{\d r^2+{r^2\over 6}\,\d\phi^2}\, .
\eea
The explicit general solution can be readily obtained. It is U-shaped and parameterized by the minimal radius $r_0$ of the profile. For any $r_0$ the asymptotic angular separation between the branches is the same $\Delta \phi={\sqrt{6}\over 4} \pi$. To understand why this is the case, we need to remember that the KW theory is conformal. The \DaD configuration breaks the conformal symmetry spontaneously by introducing the only dimensionful parameter $r_0$, while the boundary conditions at infinity $\phi\rightarrow {\rm constant}$ do not break the symmetry. As a result the mode $\partial\phi/\partial{r_0}$ is a massless Goldstone boson that does not change the boundary conditions. Besides the Goldstone boson of the conformal symmetry (and the pions -- the Goldstone bosons of the spontaneously broken chiral symmetry) all other meson modes, including the vector $\omega$, acquire a mass of order
\bea
\label{massvector}
m_{\rm mesons}\sim {r_0\over (g_s M)\alpha'}=\frac{r_0}{r_\epsilon}\, m_{\rm gb}\, ,
\eea
where $m_\text{gb}$ is the glueball mass scale
\bea
m_\text{gb} = \frac{\epsilon^{2/3}}{\lambda\alpha'}\equiv\frac{\Lambda}{\lambda}\,.
\eea
The combination $\eps^{2/3}/\alpha'$ is identified with the strong coupling scale $\Lambda$. In the regime $r_0\gg r_\eps$ the mesons are heavier than the glueballs.

In the KS case the conformal symmetry is broken explicitly by the deformation parameter $\epsilon$. In this case $r_0$ is still a free parameter $r_0\geq r_\epsilon$, but the angular separation depends on $r_0$. It is maximal $\Delta \phi=\pi$ for $r_0=r_\epsilon$ and approaches the KW value $\Delta \phi={ \sqrt{6}\over 4} \pi$ for $r_0\gg r_\eps$ \cite{DKSII}. In the regime $r_0\gg r_\epsilon$ we can treat the effect of $r_\epsilon$ as a small correction. Hence the masses of regular mesons remain essentially the same, specified by the scale $r_0$, but the massless scalar mode $\sigma$ will acquire a small mass of order
\bea
\label{massscalar}
m_{\sigma}\sim {\epsilon^{2/3}\over (g_s M)\alpha'}{\epsilon^{4/3}\over r_0^2}={r_\epsilon^2\over r_0^2}\,m_{\rm gb}\, .
\eea
We will explain this result momentarily. A priori it is not obvious why the mass of $\sigma$ should be positive, i.e. why the resulting configuration should be stable. To understand why this is indeed the case we would like to further investigate the connection between the asymptotic angle $\Delta \phi$ and $r_0$. Let us for a moment return back to the conformal case $r_\epsilon=0$ and consider a probe configuration of the following form: a straight line along the radial direction $r$ from infinity until $r_0$ at some fixed $\phi=\phi_1$, an arc at fixed $r=r_0$ between $\phi_1$ and $\phi_2=\phi_1+\Delta\phi$ and a straight line at fixed $\phi=\phi_2$  from $r_0$ to infinity. The action of such a configuration as given by (\ref{effS}) reads
\bea
\label{effSr0}
S[r_0]={r_{\rm cutoff}^4\over 2 g_s\alpha'^4}+ a r_0^4 (\Delta \phi -\Delta \phi_0)\, .
\eea
Here $a =1/(6^{1/2}g_s \alpha'^4)$ is some positive constant and $\Delta \phi_0 ={\sqrt{6}/ 2}$. Let us analyze this result.
The effective action (\ref{effSr0}) is not bounded from below if $\Delta \phi\equiv \phi_2-\phi_1 <\Delta \phi_0$. In this case the brane configuration is unstable and the U-shaped stack is pulled ``up'' towards large $r$ and disappears. The D7 and ${\overline {\rm D7}}$-branes annihilate. When $\Delta \phi >\Delta \phi_0$ the potential is bounded from below and the stack stretches down to minimize $r_0$. The resulting configuration is stable. Finally, if $\Delta\phi=\Delta \phi_0$ the parameter $r_0$ is marginal and can acquire any value. In our analysis we ignored the fact that~(\ref{effSr0}) is based on some probe configuration, which does not minimize the action even locally. Hence what we perceived as a stable configuration, minimizing $r_0$, may not in fact be such. From an explicit calculation we infer that the correct angular separation in the KW case is $\Delta \phi_0=\sqrt{6}\pi/4$. This value is larger than the approximate one (${\sqrt{6}/2}$), which is consistent with the logic that~(\ref{effSr0}) accurately predicts the fate of the unstable configurations but may not be accurate for the stable ones.

The effective action (\ref{effSr0}) has a simple field theory interpretation. The dimensionless parameter $\Delta\phi-\Delta\phi_0$ is a coupling constant in front of a marginal operator $\Phi^4$. The latter acquires a vev $\langle\Phi^4\rangle \sim r_0^4/\alpha'^4$. Indeed the non-zero value of $r_0$ is attributed on the field theory side to a vev of some chiral symmetry breaking operator~\cite{KSII}.

What happens with the effective action when we turn on $\epsilon$? Then we have different $r_0$ for different asymptotic angular  separations $\Delta \phi$. Qualitatively this can be described by the following effective action
\bea
S_{\rm eff}[\Phi,\Delta\phi]=a (\Phi^2-\Phi_0^2(\Delta\phi))^2(\Delta\phi-\Delta \phi_0)\, ,\qquad \Delta \phi_0={\sqrt{6}\over 4}\pi\, ,
\eea
and the vacuum value $\Phi_0(\Delta \phi)$ is related to $r_0(\Delta \phi)$. It is clear that for $\Delta \phi>\Delta \phi_0$ the configuration of the \DaD branes extends until $\Phi=r_0(\Delta\phi)/\alpha'$, which is a stable position with the positive mass of small fluctuation.

One should not worry about stability of the \DaD profile with respect to the $SU(2)$ charged fluctuations, i.e. the Kaluza-Klein modes on the $S^3$. This is because all such modes are massive in the KW case with the mass specified by $r_0$ as in~(\ref{massvector}). (See the details in appendix~\ref{apndx stability}.) In the $r_0\gg \epsilon^{2/3}$ regime these modes will of course remain massive.

The argument above could be made more rigorous if we reformulate the question of the sign of $m_{\sigma}^2$ as a question of stability of the classical configurations found in~\cite{DKSII}. It was explained there that the D7 action is minimized when the world-volume gauge field vanishes. As a result we can discuss stability of the U-shaped \DaD configuration in the geometry of the deformed conifold with no flux present. As was shown there the classical solution with given boundary conditions (asymptotic angular separation $\Delta \phi$) is unique and hence must be stable. Indeed there is no singularity at small $r$ for the D7-branes to end there. Also, the analysis in the KW case shows that for $\Delta \phi>\Delta \phi_0$ the D7 and $\overline{\text{D}7}$-branes do not annihilate. Hence for the given angular separation the unique solution found in \cite{DKSII} must be the minimum of the action, not just an extremum. Let us notice here that a similar argument does not work for the U-shaped configuration of the Kuperstein \DaD brane embedding~\cite{Kuperstein} (or Karch-Katz \DaD configuration in the $AdS_5\times S^5$ space~\cite{KK}). Although in that case there is a unique classical solution for the separation at  infinity, the asymptotic angular separation is zero (the branes are parallel to each other at large $r$) and the configuration is pulled to infinite $r_0$.

To calculate the  mass of $\sigma$ in terms of $r_0$ we need to know $\Delta \phi$ as a function of $r_0$ and accompany the potential (\ref{effSr0}) by a kinetic term. To estimate $\Delta \phi$ we expand the corresponding expression (it can be found in~\cite{DKSII}) at large radius. The leading deviation from the KW geometry is given by the deviation of
$$K(\tau)\sinh\tau=(\sinh\tau\cosh\tau-\tau)^{1/3}$$
from $r^2/\epsilon^{4/3}$. Expanding we get
\bea
K(\tau)\sinh\tau \simeq {r^{2}\over \epsilon^{4/3}}\left(1-{\epsilon^4\over 3 r^6}\log {2 r^3\over \epsilon^2}\right).
 \eea
As a result the leading deviation of $\Delta \phi$ from $\Delta \phi_0$ is
\bea
\Delta \phi=\Delta \phi_0- b\,{\epsilon^4\over  r_0^6}\,\log r_0\,,
\eea
where $b$ is some unimportant numerical coefficient.

Now we need to calculate the kinetic term. We substitute $\delta r_0$ for the field $\Phi$, since the geometrical meaning of the latter is the variation of $r_0$ by $\delta r_0$. To evaluate the kinetic term we simply plug the time dependent wavefunction $\phi(r,t)$ into the effective action and expand to quadratic order around the classical solution for $\phi$. In the leading order one can restrict to the zero mode solution of the conformal case, i.e. the variation $\partial/\partial r_0$ of the classical solution
\bea
\delta \phi(r)= -\delta r_0 (x_0) {\sqrt{6}\over r_0}\sqrt{r_0^8\over r^8-r_0^8}\, .
\eea
Hence the full action, up to unspecified numerical coefficients $A$ and $B$, reads
\bea
\label{actionfordeltar0}
S={1\over g_s \alpha'^4}\int \d^4x \left( A(g_s M \alpha')^2\log r_0 \left({\d \delta r_0\over \d x_0}\right)^2  - B\,  {\epsilon^4\over r_0^4}\,\log r_0\,(\delta r_0)^2\right) ,
\eea
and the mass of pseudo-Goldstone meson is given by (\ref{massscalar}).

This result has a clear interpretation in terms of an effective field theory. In the regime when $r_0\gg r_\epsilon$
we have a separation of scales. At  energies much smaller than $M_{UV}\sim \Phi={r_0/ \alpha'}$ all massive degrees of freedom associated with the D7-brane can be integrated out,
except for the light pseudo-Goldstone boson. The low-energy effective field theory is the original KS theory perturbed by some operators
\bea
\label{le}
\delta {\mathcal L}=\sum_i M_{UV}^{4-\Delta_i} O_{\Delta_i}\, .
\eea
There is also a kinetic term for the pseudo-Goldstone boson $(\partial \Phi)^2$. Comparing (\ref{le}) with (\ref{actionfordeltar0}), we conclude that the expansion starts with the operator $\epsilon^4$ of dimension $6$, i.e. it is irrelevant. The absence of relevant perturbations can be attributed to the quenched approximation, which assumes that the field theory returns to the original one once the D7-brane is completely removed ($r_0\rightarrow \infty$). This is in contrast with the pseudo-Goldstone boson of chiral symmetry in QCD, where the chiral symmetry is broken explicitly by a small mass $m_q\ll M_{UV}=\Lambda_{QCD}$
\bea
\label{smallmass}
\delta {\mathcal L}= m_q\bar \Psi \Psi\, .
\eea
Here the operator $\bar \Psi \Psi$ of dimension $3$  is relevant and therefore the effect of (2.13) on the low-energy physics increases with $\Lambda_{QCD} \rightarrow \infty$.


\section{Baryons}
\label{svi}


Following the general prescription of Witten~\cite{WittenBaryons} we introduce a baryonic vertex (b.v.) as a D3-brane wrapping the $S^3$ cycle of $T^{1,1}$. Because of the flux of the 3-form
\bea
\frac{1}{4\pi^2\alpha'}\int_{S^3}F_3 = M\,,
\eea
D3-brane carries $M$ units of charge under the $U(1)$ field on the brane. This  implies that there must be  $M$ strings ending on the wrapped D3 brane. The other ends of the strings are attached to the flavor \DaD branes, representing $M$ quarks. In this way the baryon gets flavored. The original setup~\cite{WittenBaryons} was in $AdS_5\times S^5$ which is conformal non-confining background  and the strings ended on the boundary. This is a holographic representation of a baryon made of infinitely massive quarks. A similar discussion for confining background was presented in~\cite{Brandhuber:1998xy,Callan:1999zf}. Dynamical baryons, as opposed to infinitely massive ones, were introduced by ending the strings on flavor branes rather than on the boundary. This was done in the context of the Sakai-Sugimoto model in~\cite{Bergman:2007wp,Seki2008}.

The strings pull the b.v.  towards  the \DaD\!\! branes. This is opposed by the geometry which, through the DBI action, attracts the b.v. to the tip of the conifold. In the case of the Sakai Sugimoto  model the  strings always win~\cite{Seki2008}. As we are going to demonstrate, this is not the case for the KS background.

The energy of the b.v.  at a point $r=\epsilon^{2/3}\cosh^{1/3}\tau$ is given by (here we use the notations of \cite{KS,remarks})
\bea
\frac{1}{{\alpha'}^2}\int_{S^3}e^{-\Phi}\sqrt{-g_{D3}}\sim  \frac{\epsilon^2}{g_s{\alpha'}^2}\, h^{1/2}(\tau)\cosh\tau\,.
\eea
Here $h(\tau)$ is the warp-factor of the KS geometry. For large $r$ the energy scales as
\bea
E_{\text{b.v.}} \sim \frac{M}{\alpha'}\, r\log\frac{r}{r_\epsilon}\,.
\eea

The energy of  the $M$ strings is given by
\bea
E_\text{strings}=M\frac{1}{\alpha'}\int_r^{r_0} \sqrt {- g_{\tau\tau}g_{00}}\,\d r \sim \frac{M}{\alpha'}\int_\tau^{\tau_0}\frac{\epsilon^{2/3}}{K(\tau)}\, \d \tau\, ,
\eea
which for large $r$ behaves as
\bea
\label{f1}
E_\text{strings} \sim \frac{M}{\alpha'}\,(r_0-r)\,.
\eea

{
\begin{figure}[htb]
\begin{minipage}[b]{0.5\linewidth}
\begin{center}

\includegraphics[width=6.cm]{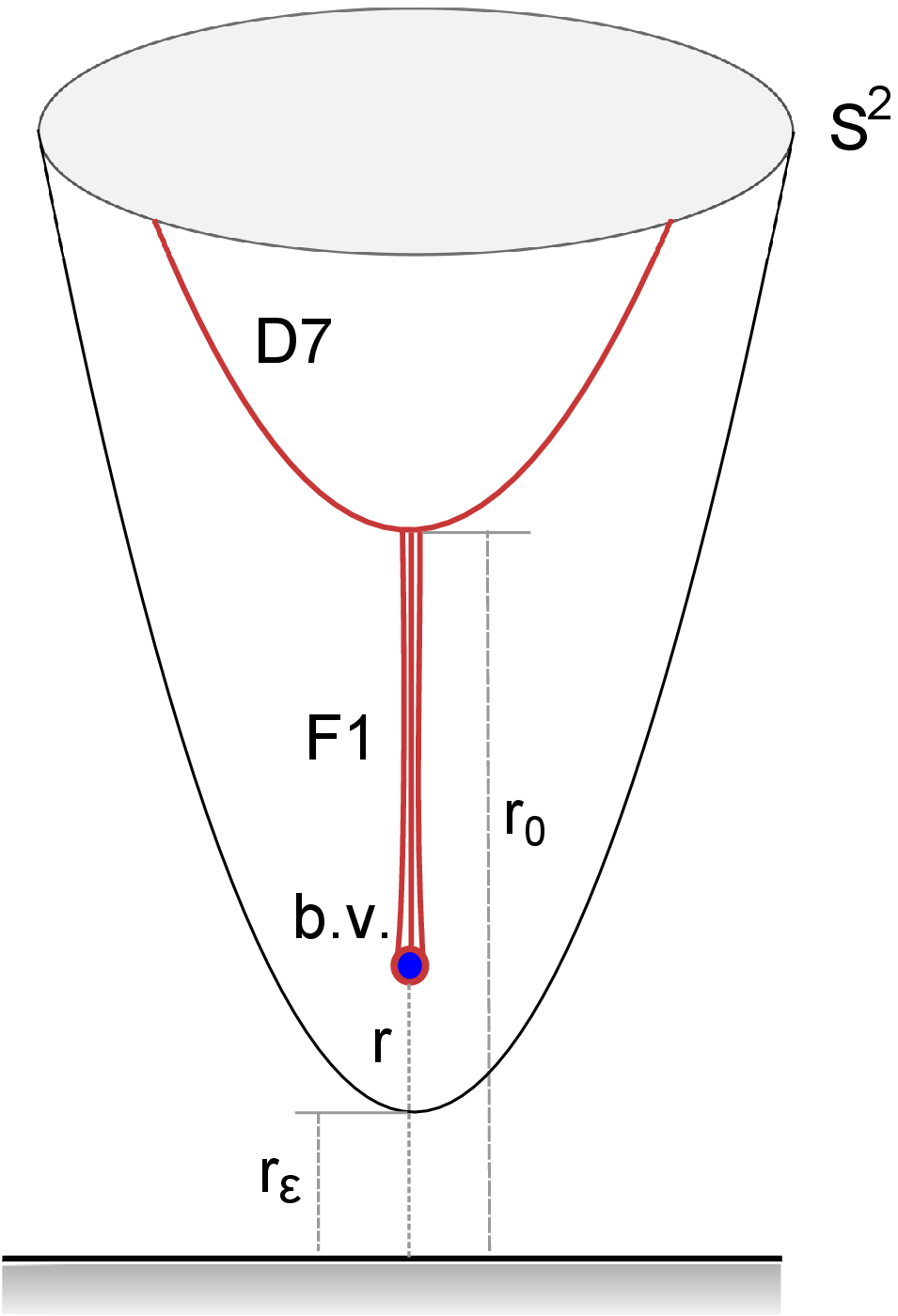}

(a)
\end{center}
\end{minipage}
\begin{minipage}[b]{0.5\linewidth}
\begin{center}

\includegraphics[width=6.cm]{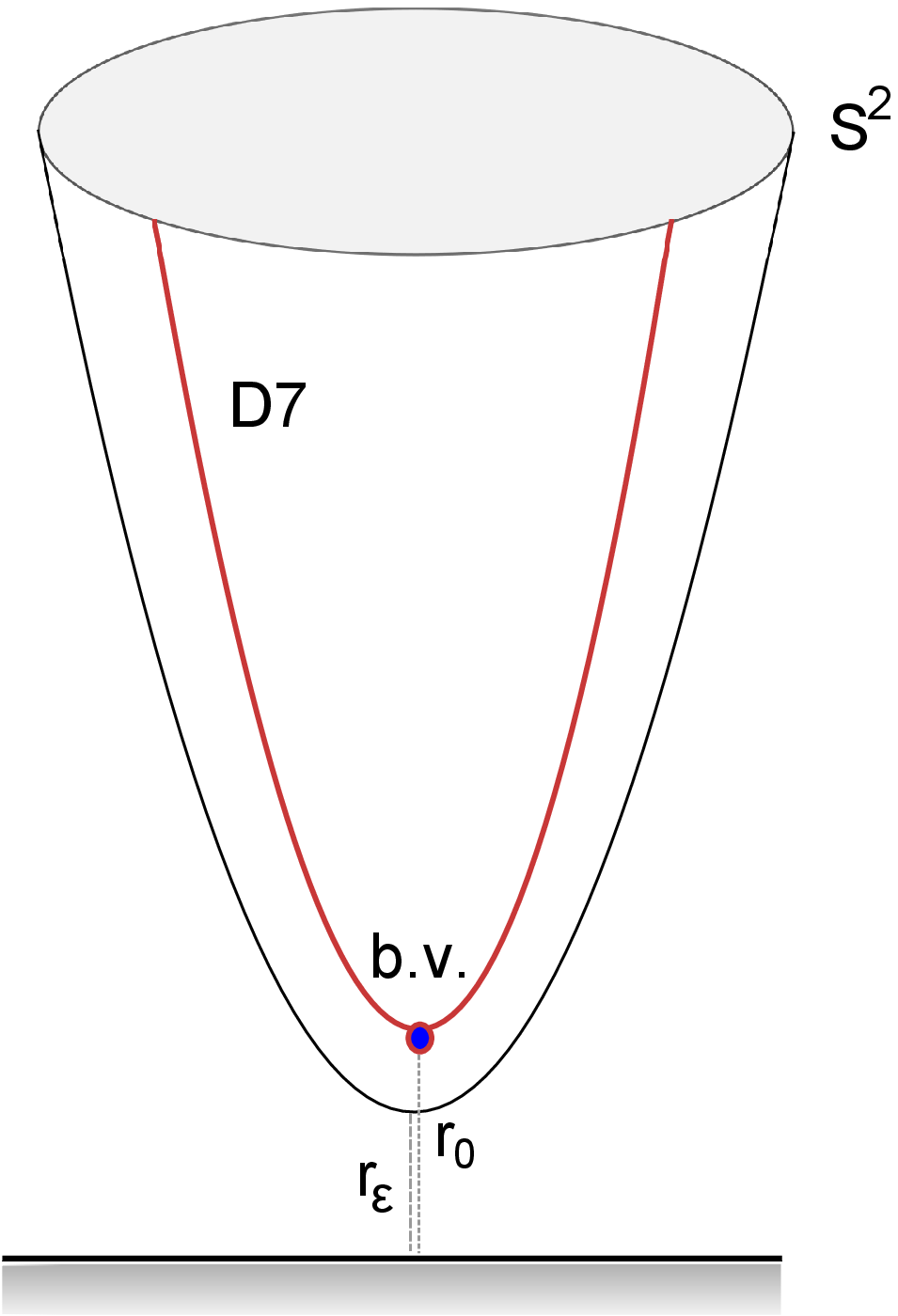}

(b)
\end{center}
\end{minipage}
\vspace{-0.6cm}
\caption{\small (a) Baryon as a D3-brane wrapped on the $S^3$ of the conifold. It sits between the tip of the conifold ($r_\eps$) and the lowest point of the \DaD configuration ($r_0$). $M$ strings, stretched between the D3 and the D7, pull the D3 up, but the geometry force is stronger and keeps the D3 close to the tip of the conifold. (b) If the \DaD configuration descends close enough to the tip, the string tension wins over the gravity and the D3 dissolves in the D7 becoming a flavor instanton.}
\label{fig svi}
\end{figure}}

The position of the D3 along the radial direction will be determined  by minimizing  the total  energy
\bea
E_{\text{tot}} = \frac{M}{\alpha'}\left(A\, r\log r + B |r_0-r|)\right),
\eea
where $A$ and $B$ are some numerical coefficients. Hence the b.v. will sit at $r\sim r_{\text{IR}}$. Keeping in mind that there are $M$ strings which can affect each other through backreaction, it is hard to find the  actual values of $A$ and $B$. Nevertheless, it is clear that for large $r_0$ the D3 will prefer to settle in the IR and the baryon will correspond to $M$ strings stretching between the D3 and \DaD (figure~\ref{fig svi}).

In such a case the properties of the baryon are determined by a composite system of the b.v., the $M$ strings, their  endpoints  on the b.v.,  and their flavored endpoints on the \DaD branes. The endpoints on the b.v. are fermionic thus assuring the antisymmetric product of $M$ fundamentals~\cite{WittenBaryons}. On the other hand the flavored endpoints have to be bosons so that one can recast using the Hartree approximation the well known picture of large $M$ baryons~\cite{WittenlMB}. In this picture the size of the baryon scales with $M^0$ and the mass as $M$.

For small $r_0$ the space is locally flat, the energy of the wrapped D3 will be approximately independent of the location $r$ and the strings can win. Expanding the above integrals around $\tau=0$, one finds
\bea
\label{massofbaryon0}
E_{\text{b.v}}\sim \frac{M}{\alpha'}\, \epsilon^{2/3} \left(I^{1/2}(0) + O(\tau^2)\right) ,
\eea
where the warp factor was traded for a dimensionless function $h\propto (g_sM\alpha')^2\eps^{-8/3}I(\tau)$, and
\bea
E_\text{strings} \sim \frac{M}{\alpha'}\, \epsilon^{2/3}\left(|\tau_0-\tau| + O(\tau_0^{5/3})\right).
\eea
In the region of sufficiently small $r_0$, in which one can approximate the energy of D3 by a constant, D3 will be pulled towards the \DaD and dissolve in them, i.e. it will be equivalent to an instanton configuration of the gauge fields on the \DaD world-volume~\cite{Dbrane-instanton}. In the latter case the instanton coupling comes from the Chern-Simons action of the \DaD branes:
\bea
\label{CScoupling}
S_{\text{CS}}=\frac{1}{96\pi^4{\alpha'}}\int C_2\wedge\Tr\F^3 =\frac{1}{96\pi^4{\alpha'}}\int F_3\wedge\omega_5({\cal{A}})\, ,
\eea
where $\F = \d \A + \A\wedge \A$ and
\bea
\label{cs}
\omega_5({\cal{A}})= \Tr\left(- \A\wedge \F^2 + \frac{1}{2}\,\A^3\wedge \F + \frac{19}{10}\,\A^5\right).
\eea
Expanding the $U(N_f)$ field into the $SU(N_f)$ and the $U(1)$ parts
\bea
\label{UNtoSUN}
{\cal{A}}= A + \frac{1}{\sqrt{2N_f}}\, \hat{A}\cdot{\mathbb{I}}_{N_f}\,,
\eea
one obtains from the first term in (\ref{cs})
\bea
\label{CS eff}
S_{\text{CS}}\sim M\int \d^4x\,\d r\, \hat{A}_0\,\Tr F_{mn}^2\,.
\eea
$F_{mn}$ is the field strength associated with the $SU(N_f)$ gauge fields and the indices $m,n=1,2,3,4$ run through the space coordinates and the radial direction. For the instanton configuration of the non-abelian field $A$ the result is the instanton number $n$ times $M$ units of charge under the $U(1)$ gauge field $\hat{A}$. Since the field $\hat{A}_0$ is dual to the baryon number chemical potential in the dual field theory, $n$ is the baryon number.

The effective 5-dimensional action for the gauge fields on the \DaD branes consists of the interaction part~(\ref{CScoupling}) and a kinetic term, obtained from the expansion of the DBI action~(\ref{DBI}):
\bea
\label{DBI eff}
S_\text{DBI}= - \frac{1}{4 g_s{\alpha'}^2}\int \d^4 x\, \d z\, k(z)\, \Tr\big[ h(z) \F_{\mu\nu}^2+2 \F_{\mu z}^2\big]\, .
\eea
Here a new radial coordinate $z$ is introduced, such that
\bea
\label{zmetric}
\d s_8^2 = h^{-1/2}\d s_{3,1}^2 + h^{1/2}\left(\d z^2 + r^2(z)\d\Omega_3^2\right)
\eea
and
\bea
\label{k(z)}
k(z)=\frac{\epsilon^2\cosh\tau(z)}{16\sqrt{2}\pi^3}\,.
\eea
Integration  over $z$ in (\ref{DBI eff}) is implied over both branches of the \DaD branes. The solutions to the equations of motion can be found as an expansion in $\lambda^{-1}$, starting from flat-space BPS instantons as a leading order approximation.

In this approximation the mass of the instanton is given by the expression~(\ref{massofbaryon0}):
\bea
m_{\text{baryon}}\sim \frac{M}{\alpha'}\,\eps^{2/3}\equiv M\Lambda\,.
\eea
Higher order corrections come from the non-zero curvature at the position of the baryon and the Chern-Simons interaction. A more detailed analysis of corrections is done in appendix~\ref{sizeofbaryon}. It is very similar to the analysis in the Sakai-Sugimoto setup~\cite{SSb}. Here we notice that also in the KS case the size of the baryon remains small,
\bea
\label{size}
\rho \sim {\epsilon^{2/3}\over \lambda^{1/2} }\, ,
\eea
which makes higher string corrections important.


\section{Glueball induced potential}
\label{glueballs}
Now we turn to the interactions between baryons. First we will estimate the glueball contribution to the baryon-baryon interaction potential. To calculate the potential we would need to know how  glueballs couple to baryons. Notice first that the scalar and vector glueballs couple to the baryon in different ways. The former couples through the DBI term, while the latter through the Chern-Simons term. Let us start with the scalar. We will consider a mode which warps the six-dimensional geometry of the conifold
\bea
\label{xmetric}
\d s^2=h^{-1/2}\d s^2_{3,1} + \text{e}^X h^{1/2} \d \tilde{s}_6^2\, .
\eea
Here $X$ is a small perturbation and we are going to find the effective action describing its dynamics. In the KS background this mode apparently couples to other glueballs. We will neglect this interaction as it will only change the numerical coefficient, but not the parameter dependence of the interaction. We start with the action in the bulk expanded to the second order in $X$:
\bea
\label{variation bulk}
S_{\rm bulk}\propto \frac{1}{g_s^2\alpha'^4} \int \d^{10}x \,\delta g^{AB} \delta R_{AB}+...
\eea
We focus on the singlets with respect to the $SU(2)\times SU(2)$ symmetry as the charged glueballs tend to be more massive. Now we use the relation
\bea
\delta R_{ij}={1\over 4}\,\tilde{g}_{ij} h\, \partial^2 X\, ,
\eea
where $\partial^2$ is the Minkowski Laplacian. After integration by parts and evaluating the integral along the radial $\tau$ and angular directions  we arrive at the effective action
\bea
S_{\rm bulk}[X]\to\left(M{\epsilon^{2/3}\over \alpha'}\right)^2\int \d^4x \left((\partial \tilde{X})^2-m_4^2\, \tilde{X}^2\right) .
\eea
Here we separated the Minkowski and $\tau$ dependence of the wave function $X(x^\mu,\tau)=\tilde{X}(x^\mu)\psi(\tau)$. The four-dimensional mass $m_4^2$ has a typical scale of the glueball mass $m_{\rm gb}$. Its numerical values were computed in \cite{Berg}. It was also shown there that there are modes with $\psi(\tau)$ not vanishing at the tip $\psi(0)\sim 1$.

The interaction with the baryon is even easier to find. As we discussed earlier the baryon can either be described by a D3-brane wrapping the $S^3$ cycle or by an instanton inside the \DaD flavor brane. In any regime it sits close to the tip of the conifold $r\sim r_\epsilon$ and its mass is given (at the leading order) by $M{\epsilon^{2/3}/ \alpha'}$. The warp mode $X$ will change the geometric volume of $S^3$ and hence the mass of the baryon, leading to the interaction term
\bea
S_{\rm D3}[X]=\left(M{\epsilon^{2/3}\over \alpha'}\right) \int \d x^0\, \tilde{X} \psi(0)\, .
\eea
Taking $\psi(0)\sim 1$ we find the glueball induced potential for two baryons separated by the distance $|x|$ to be
\bea
\label{glueball potnl}
V_{\rm gb}\sim {\text{e}^{-m_{\rm gb}|x|}\over |x|}\, .
\eea

The derivation of the repulsive vector meson potential proceeds along the same lines. Vector glueballs even under charge conjugation were studied in~\cite{Gordeli}. Contrary to the scalar glueballs, vector equations can be diagonalized and one can enjoy using the explicit form of the wavefunction. Of particular interest is the lightest vector dual to the $U(1)_R$ current, which appears from the fluctuations of the metric and 5-form field. This glueball interacts with the baryon through the coupling of the wrapped D3 brane to the four form $C_4$:
\bea
S_{\text{D3}} \sim  \frac{1}{{\alpha'}^2}\int_{\text{D3}} \delta C_4\, ,
\eea
where $\delta C_4$ can be read off from~\cite{Gordeli}:
\bea
\delta C_4 \sim g_s M^2{\alpha'}^2\, V\wedge g_5\wedge \d g_5+ \ldots , \qquad g_5 = \d \psi + \dots
\eea
Here $V=\varphi(\tau)\tilde{V}_\mu(x^\nu)\, \d x^\mu $ denotes the bulk glueball field and the conventional para\-met\-ri\-zation of $T^{1,1}$ is used~\cite{KS}. The fluctuation also affects the metric
\bea
\delta(\d s^2) \sim \frac{(g_s M \alpha')^2}{K\sinh\tau}\, V\cdot g^5\,.
\eea
Plugging this into equation~(\ref{variation bulk}) we arrive at the following expression for the effective action for $\tilde{V}_\mu$
\bea
S_{\text{bulk}}[\tilde{V}] \to g_s^2 M^4 \int \d^4 x \left((\partial_\mu \tilde{V}_\nu)^2 - m_4^2 \tilde{V}_\mu^2\right) ,
\eea
which gives the same interaction potential~(\ref{glueball potnl}) modulo a numerical prefactor.

As we will infer from the next section, the interaction carried by mesons is enhanced by $g_s^{-1}$ as compared with the one by glueballs. Therefore we will neglect glueballs in the later discussion. Notice however that smallness of $g_s$ is dictated by the holographic approach.

So far we have ignored the massless glueballs of the KS theory~{\cite{Aharony:2000pp,GHK}, which could dominate the potential at large distances. There are two exactly massless particles: the Goldstone boson of the spontaneously broken $U(1)_{B}$ baryon symmetry and its CPT-partner. These glueballs are odd with respect to the $\mathcal I$-symmetry, which we have discussed above. Since the geometry of the deformed conifold is invariant under this symmetry the $S^3$ cycle of minimal volume located at the tip of the conifold is also invariant. As a result the baryon wrapping this cycle does not couple to the $\mathcal I$-odd massless glueballs at the lowest order. The $\mathcal I$-symmetry is broken by the presence of \DaD branes. The D3$-$D7 strings would pull the D3-brane in a $\mathcal I$ non-invariant location. This in turn would create a coupling between the baryon and massless glueballs. The wavefunction of these glueballs behaves as $\tau^2$ near the tip~\cite{GHK} and provided that the D3-brane is sufficiently close to the tip the coupling is not large. Although these glueballs will be responsible for the dominant force at large distances their contribution to the binding energy will still be $g_s$ suppressed compared with $\sigma$ and $w$ mesons. This argument allows us to neglect the effect of these particles making our model more similar to hadron physics.


\section{Meson induced potential}
\label{mesons}

\subsection{Regime of large $r_0\gg  r_\epsilon$}
\label{regime1section}
In this regime the baryon is the b.v. that sits close to the tip of the conifold and is connected with $M$ strings to the \DaD branes. The pseudo-Goldstone boson $\sigma$ is the lightest particle except for the massless pions (and $\eta'$) and the pair of massless glueballs. This fact guarantees the attraction of two baryons at large distances. To estimate the binding energy we need to combine the action~(\ref{actionfordeltar0}) for $\sigma$ with the term describing its interaction with the baryon through the D3$-$D7 strings.

When $r_0$ is large the pseudo-Goldstone boson is simply generated by the dilatation $\partial/\partial r_0$. Hence to find the interaction term we need to estimate how the energy of the baryon (and the attached strings) will change when $r_0$ is changed. Clearly the main contribution comes from the energy of the D3$-$D7 strings, which depends on $r_0$~(\ref{f1}). Hence the interaction term is
\bea
{M\over \alpha'}\int \d x^0\, \delta r_0 \, .
\eea
Together with (\ref{actionfordeltar0}) this gives the attractive potential
\bea
\label{meson potnl1}
V_{\rm scalar}\sim -{1\over g_s \log r_0}{\text{e}^{-m_{\sigma}|x|}\over |x|}\ .
\eea
As was mentioned before the potential energy is $1/g_s$ times enhanced as compared with that of the glueballs.

The vector meson couples to the strings as follows. The endpoints of the strings act as sources for the gauge fields on the \DaD world-volume. For a static baryon we obtain in the point-source approximation
\bea
S_{\text{string}}[\hat{A}]= M\int \d^5 x\ J^0 \hat{A}_0 \sim M \int \d x^0\ \hat{A}_0\,.
\eea
Alternatively this result can be obtained from the boundary term in the action for the strings:
\bea
M \int \hat{A}\, .
\eea

The kinetic term for $\hat{A}_0=\hat{a}_0(x^\mu)\varphi(r)$ comes from the DBI action on the \DaD\!:
\bea
S_{\text{DBI}}[\hat{a}] \sim g_s M^2 \int \d^4 x\left((\partial \hat{a}_0)^2-m_{\omega}^2 \hat{a}_0^2\right).
\eea
This gives the same expression~(\ref{meson potnl1}) for the potential modulo the sign and a numerical prefactor.

A more detailed analysis is required in order to understand, whether the repulsion will overcome the attraction at short distances. Such a calculation however is beyond the scope of the present paper.
A more interesting prospect opens in the regime of small $r_0$, which we discuss in the next section.


\subsection{Regime of small $r_0\sim r_\epsilon$}
\label{regime2section}
In this regime the D3-brane dissolves inside the \DaD to form an instanton. To calculate the meson induced potential at large distances, we need to calculate the interaction between the mesons and the instanton solution on the D7-branes. We start with the pseudo-Goldstone boson, which in this case has a mass of order $m_{\sigma}\sim m_{\rm gb}$. Its wavefunction can still be parameterized by the deviation of the lowest point of the profile $\delta r_0$. But in this case there is only one scale $r_0\sim r_\epsilon$ and it is actually easier to estimate the result than in the $r_0\gg r_\epsilon$ regime. The wavefunction can be expanded into a product of the Minkowski wave function $\sigma(x_\mu)$ and the internal wavefunction $\psi(\phi)$, which depends on the angular variable along the $S^2$. (We choose $\phi$ instead of the radius because it is a good coordinate near the tip $\phi=0$.) The internal wavefunction is of  the order of $\psi(0)\sim r_\epsilon$ and a simple estimate gives, cf.~(\ref{actionfordeltar0}),
\bea
S_{\rm eff}[\sigma]\sim {(g_s M \alpha' r_\epsilon)^2\over g_s \alpha'^4}\int \d^4x \left(   \left(\partial \sigma\right)^2 -m_{\sigma}^2 \sigma^2\right) .
\eea
To calculate the interaction term we need to expand the DBI action keeping the dependence on the non-abelian gauge fields,
\bea
\label{na}
S_{\rm na}\sim {1\over g_s {\alpha'}^2}\int \d^4 x\,\d\phi\ r^3\sqrt{r^2+6 (r')^2}\,\Tr\left(hF_{ij}^2+\frac{12F_{i\phi}^2 }{r^2+6 (r')^2} \right),
\eea
and then expand up to the lowest order in $\delta r$. In the above expression $r'\equiv {\d r/ \d\phi}$. Taking into account that the non-abelian fields present a point-like instanton at the origin a straightforward calculation gives
\bea
\label{na2}
S_{\rm na}\sim {g_s M \alpha' r_\epsilon \over g_s \alpha'^2}\int \d x^0\,  \sigma\, .
\eea
Therefore the resulting potential is
\bea
\label{for}
V_{\rm scalar}= - {g_\sigma^2\over 4\pi g_s}\, {\text{e}^{-m_{\sigma} |x|}\over |x|}\, .
\eea
Here we have introduced a dimensionless coupling constant ${g_\sigma^2/ (4\pi)}$ to make a connection with the nuclear physics based models, e.g.~\cite{Walecka}. The precise value of this coupling is not important to us. We know however that it has to vanish when $r_0$ approaches $r_\epsilon$. This is because the case $r_0=r_\epsilon$ is a special antipodal configuration, for which the world-line of \DaD on $S^2$ degenerates to two antipodal points.  In terms of parametrization $\phi=\phi(r)$, the $\Z_2$ inside the emergent $U(1)$ would just flip the sign of the fluctuation $\delta\phi$. Therefore in the leading order $\delta\phi$ cannot couple to the baryon and $g_\sigma$ must vanish. This is not obvious from (\ref{na}), because the expressions we used are for the KW case, which is not a good approximation for small $r_0$. Our justification of using them to calculate the potential~(\ref{for}) in the regime of small $r_0$ is that when $r_0$ is small but not very close to $r_\epsilon$ the KW approximation would give a correct answer up to an overall factor  of order one. To see how the coupling constant vanishes in the antipodal case we switch to the KS coordinate $\tau(\phi)$, where $r^3=\epsilon^2 \cosh\tau$, and remind the reader the form of the induced metric near the tip $\tau=0$~\cite{DKSII}:
\bea
\d s^2=h^{-1/2}\,\d s_{3,1}^2+g_{\tau\tau}\, \d\tau^2+g_{\phi\phi}\, \d\phi^2+\sum_{i=1}^3 g_{ii}\,e_i^2\, .
\eea
The functions $h$, $g_{\tau\tau}$ and $g_{ii}$ expand as $c_0+c_1\tau^2$ at small $\tau$, while $g_{\phi\phi}\sim \tau^2$. Then the coefficient in the interaction term (\ref{na2}) becomes proportional to
\bea
\label{efftau}
\int \d\phi \ \sqrt{\zeta+{\tau'}^2 \tau^{-2}} \,\delta(\phi)\,,
\eea
where $\zeta$ is some numerical constant and $\delta(\phi)$ is the delta-function. The geometry of almost antipodal configuration can be described as follows:  $r$ as a function of $\phi$ changes very rapidly from infinity to $r_\epsilon$ in a very short region near $\phi=-{\pi/ 2}$; then it stays constant until $\phi$ almost reaches ${\pi/2}$ and after rapidly goes to infinity. This means that $\tau$ is small and constant for almost all values of $\phi$. Near the tip $\phi\rightarrow 0$, $\tau$ approaches zero, but $\tau'$ approaches zero much faster. Therefore expanding (\ref{efftau}) up to the linear order in $\delta \tau$ would not result in any non-zero coupling.

Similarly we calculate the potential induced by the exchange of the vector mesons. Action for the electric part of the $U(1)$ field~(\ref{UNtoSUN}) on the \DaD branes reads
\begin{multline}
S_{\rm eff}[\hat{A}]\sim {\alpha'^2 \over g_s \alpha'^2}\int \d^4 x\, \d\phi\ r^3\sqrt{r^2+6 (r')^2}\left(h (\partial_\mu \hat{A}_0)^2-\ldots\right)
\\  \to g_s M^2 \int \d^4 x\left((\partial_\mu \hat{a}_0)^2-m_{\omega}^2 \hat{a}_0^2\right) ,
\end{multline}
where the dimensionless $\hat{A}_0(x^\mu)$ was expanded in terms of dimensionless functions $\hat{a}_0(x^\mu)$ and $\varphi(\phi)$. The interaction term comes from the Chern-Simons part of the action
\bea
\label{CSinteraction}
S_{\rm CS} \sim M \int \d^4 x\, \d\phi\  \hat{A}_0 \frac{\epsilon^{ijk}}{2}\,\Tr(F_{ij}F_{k\phi})\sim M \int \d^4 x \,\hat{a}_0 \varphi(0)\, ,
\eea
with $\varphi(0)\sim 1$. The resulting potential is simply
\bea
V_{\rm vector}=\frac{g_\omega^2}{4\pi g_s}\,{\text{e}^{-m_{\omega} |x|}\over |x|}\, .
\eea

So far we have ignored the pions, the Goldstone bosons of the broken chiral symmetry. These particles are massless and, provided their coupling to baryons is not suppressed, the one-pion exchange (OPE) would give a dominant contribution. Thus the processes involving OPE are important in the case of interactions of individual baryons. The effective pion-nucleon vertex has the form
\bea
\label{pionvertex}
 2i g_A \,\frac{\partial_\mu  \pi^a }{f_\pi}\,\overline N(p') \gamma_5 \gamma^\mu \tau^a N(p)\, .
\eea
In the case of massless pions the resulting potential is
\bea
\label{vpi}
V_{\pi}={g_A^2\vec{\tau_1}\cdot\vec{\tau_2}\over 4\pi f_\pi^2}\,{S_{12}\over |x|^3}\ .
\eea
Here $\tau_i$ is the operator of isospin and the dependence on the operator of spin $\sigma_i$ enters
\bea
S_{12}=3{(\vec{x}\cdot \vec{\sigma_1})(\vec{x}\cdot \vec{\sigma_2})\over x^2}-\vec{\sigma_1}\cdot\vec{\sigma_2}\,.
\eea

A similar potential appears in the Sakai-Sugimoto model~\cite{Hashimoto}. In that case the low-energy baryon-pion interaction can also be described by the coupling~(\ref{pionvertex}) with appropriate $g_A$ and $f_\pi$. The pions $\pi$ and massless isoscalar meson $\eta'$ come from the fluctuations of the gauge field
\bea
\label{pion}
{A}_\mu=U^+ \partial_\mu U \,\psi_+(\phi)\, ,\qquad U=\exp\left({i(\eta'+\pi) \over f_\pi}\right) , \qquad \partial_\phi \psi(\phi)=(\sqrt{g}g^{\phi\phi})^{-1}\, .
\eea
However the expansion of the D7-brane action around the classical solution for the static baryon does not lead to an interaction term. To obtain~(\ref{pionvertex}) and calculate $g_A$ one has to quantize the baryons~\cite{Hashimoto:2008zw}. The resulting potential, besides~(\ref{vpi}), also includes an extra term due to the exchange of $\eta'$. The latter coincides with~(\ref{vpi}) modulo the isospin interaction term $\vec{\tau_1}\cdot\vec{\tau_2}$ and the value of $g_A$.

We expect our model to be similar to the Sakai-Sugimoto model. Namely we expect the pion (and $\eta'$) potential to be of the form~(\ref{vpi}) with appropriate values of coefficients and isospin structure. Although we do not compute $g_A$ and $f_\pi$ in this work, we expect them to satisfy
\bea
{g_A^2\over f_\pi^2}\sim {1\over g_s m_{\rm gb}^2}\, .
\eea
Clearly the massless pion tail will dominate the large distance potential. At the characteristic scale of the $\sigma$ and $\omega$ potentials, that is $x\sim m_{\rm gb}^{-1}$, the OPE potential will be of the same order as the $\sigma$ and $\omega$ contributions.

It is believed however that for heavy nuclei or baryonic matter the contribution of the OPE is largely averaged to zero due to the tensor structure. Following the nuclear physics models of such systems we will ignore its effect on the potential. Since the two (or more) particle exchanges correspond to loop diagrams and therefore $g_s$ suppressed, we are ready to proceed to a discussion of our findings: in the next section we will present a holographic model of weakly bound baryons.


\section{Conclusions. A holographic model of baryons}
\label{conclusions}

In this paper we considered a holographic model of baryons based on non-super\-sym\-metric D7 and $\overline{\text{D7}}$-branes embedded in the KS geometry. We addressed the question of baryon interactions and in particular searched for a model of baryons that attract at large distances and repel at small. A more specific goal was to shed some light on the problem of smallness of the binding energy. These are the identifying properties of baryons.

The holographic approach gave us control over the meson dynamics and made other contributions subleading. We have observed that the holographic limit $\lambda\to~\!\infty$, $g_s\to 0$ suppressed the glueball contribution to the binding energy, as well as contribution of the multi-particle exchange. We have also discarded the contribution of massless pions by assuming the case of heavy nuclei, or more generally, baryonic matter. This leaves us with the lightest massive meson exchanges as the leading interactions of baryons. The two competing forces are the attraction induced by the meson $\sigma$ which corresponds to fluctuations of the embedding profile and the repulsion due to the meson $\omega$ described by the fluctuation of the $U(1)$ gauge field on the world volume of the \DaD branes.

In the model we considered, the masses of $\sigma$ and $\omega$ are parameterized by $r_0$, the position of the lowest point of the \DaD embedding. For values of $r_0$ far from the IR end of the geometry $r_\eps$, $m_\sigma\ll m_\omega$. This ensures the attraction of baryons at large distances. The analysis of this paper does not allow to conclude whether the attraction would switch for repulsion at small distances, but even if it does so there is no natural suppression for the binding energy except for the small $\lambda^{-2}$ factor specific to holographic models.

A more interesting regime is $r_0\sim r_\eps$, when $m_\sigma\sim m_\omega$. This paper does not give the answer, whether $m_\sigma<m_\omega$ always holds, but one can certainly tune the parameter $r_0$ to satisfy this condition. This sounds promising because $m_\sigma\sim m_\omega$ itself implies the near-cancelation of the forces. The large distance potential has a typical form for effective models of nuclear interactions, e.g.~\cite{Walecka}:
\bea
\label{conclpotnl}
V = \frac{1}{4\pi g_s}\left( g_\omega^2\,{\text{e}^{-m_{\omega} |x|}\over |x|} - g_\sigma^2\, {\text{e}^{-m_{\sigma} |x|}\over |x|} \right).
\eea
For $m_\sigma<m_\omega$ and $g_\sigma<g_\omega$ the potential is attractive at large and repulsive at small distances. As we discuss in detail in the text, in the regime $r_0\sim r_\eps$ the coupling constant $g_\sigma$ should be small. This is due to an emergent $\mathbb{Z}_2$ symmetry, which forces $g_\sigma$ to vanish in the case $r_0=r_\eps$. Therefore one can tune $r_0$ to satisfy $g_\sigma<g_\omega$. A crucial question which needs further investigation is whether one can tune $r_0$ to satisfy both $g_\sigma<g_\omega$ and $m_\sigma<m_\omega$ simultaneously.

Assuming the answer is positive the resulting model will share the basic property of the nucleon physics. The binding energy of two baryons will be suppressed by  the similarity of the masses of the mesons and potentially by the smallness of the ratio $g_\sigma^2/g_\omega^2$.

Is this a quantitative model of the baryonic interaction? First of all, one has to match the mesons $\sigma$ and $\omega$ to the real world mesons. This problem and in particular, the fate of $\sigma$ was extensively discussed in~\cite{Kaplunovsky}. Next, one can try to match the masses and couplings in the potential~(\ref{conclpotnl}) to the real world values. In the present model all parameters, including mass ratio and the value of couplings depend on a single parameter $r_0$. It seems there is no reason for the matching. Finally, one will have to say something about the contribution of pions, multi-particle exchanges and glueballs, once the holographic limit is relaxed and they are no longer suppressed.

\bigskip
To summarize, we formulate the two key ingredients that give us a hope to build holographic models of baryon interactions. First, for attractive baryons, one has to satisfy $m_\sigma<m_\omega$. In our case we relied on the conformal symmetry in the UV. The meson $\sigma$ is a pseudo-Goldstone boson and hence it is natural that it is light. Second, one has to satisfy the condition $g_\sigma<g_\omega$. In both the Sakai-Sugimoto and the present model this can be achieved by dialing $r_0$. At smallest value of $r_0$ there is an emergent $\mathbb{Z}_2$ symmetry which forces the coupling to vanish. The complication may arise while combining these two ingredients. The light pseudo-Goldstone boson requires the breaking scale $r_0$ to be much larger than the internal scale of the gauge theory $r_\epsilon$. At the same time the $\mathbb{Z}_2$ symmetry emerges when $r_0=r_\epsilon$. The next crucial step would be to calculate the masses of $\sigma$ and $\omega$ for a general configuration to ensure $m_\sigma<m_\omega$. We leave this task for a future work.


\vspace{0.5cm}
\centerline{\bf Acknowledgments}
\vspace{0.2cm}
We thank O.~Aharony, T.~Banks, O.~Bergman, G.~Festuccia, V.~Kaplunovsky, I.~Klebanov, Z.~Komargodski, S.~Kuperstein, J.~Maldacena, S.~Nussinov, V.~Pestun, S-J.~Rey, M.~Strassler, A.~Vainshtein, M.~Voloshin, S.~Yankielowicz and I.~Zahed for useful discussions. A.D. thanks the theory group at Tel-Aviv University for hospitality while this work initiated. D.M. appreciates hospitality of the NHETC at Rutgers, FTPI at the University of Minnesota and the IAS at Princeton, where a part of this work was done. The work was partially supported from the following sources: the DOE grant DE-FG02-90ER40542, the Monell Foundation, the RFBR grant 07-02-00878 (A.D); the RFBR grant 07-02-01161 and the contract 14.740.11.0081 with the Ministry
of Education and Science of the Russian Federation (D.M.); the Grant for Support of Scientific Schools NSh-3035.2008.2 (A.D. and D.M.); the grant 1468/06 of the Israel Science Foundation, the  grant (DIP H52) of the German-Israel Project Cooperation, a grant of the German Israeli foundation  and  a grant of  the US-Israel Binational Science Foundation (D.M. and J.S.).


\begin{appendix}

\section{Stability of the \DaD embedding}
\label{apndx stability}
To show local stability of some classical solution one has to show that the spectrum of small perturbations is positively defined. Although simple conceptually, this task could be complicated in practice. In the case of interest the main problem would come from the interaction of fluctuations of geometrical shape with the world-volume gauge fields. Instead we would rather prove stability by showing that the energy (action for the static configuration in question) reaches its minimum on the solution of interest. It was shown in~\cite{DKSII} that the action of the D7-brane is calibrated by the geometrical volume it spans
\bea
\label{stabfunct}
S_\text{D7}=S_{\text{DBI}}+S_{\text{CS}}\geq {\rm Vol}_4\int \d^4y\,\sqrt{g}\, .
\eea
Therefore we are going to show that the volume can only increase should there be a small fluctuation around the original solution. This logic allows us to disregard the dynamics of the world-volume gauge fields and from now on we simply put them to zero. Let us point out that while the world-volume gauge field does not affect stability, the effect on stability of the fields in the Minkowski directions is far from being trivial~\cite{Ooguri:2010xs}.

In general our goal should be stability of the \DaD configuration in the KS case. However the corresponding calculation would be quite complicated. Instead we will focus on the KW case. When $r_0\gg r_\epsilon$ the geometry of the KS solution approaches the one of the KW solution. We will find that except for the massless Goldstone boson $\sigma$ all other modes governing the shape are massive. It is argued in section~\ref{massesofmesons} that the Goldstone boson acquires some positive mass when we turn on a non-zero $\epsilon$. This at least proves stability in the regime $r_0\gg r_\epsilon$.

When $r_0\sim r_\epsilon$ the KW analysis is not valid. But unless the D7-branes tend to spontaneously break the $SU(2)$ symmetry of the KS solution stability follows from the uniqueness of the $SU(2)$ invariant classical solution found in~\cite{DKSII} and the observation in section \ref{massesofmesons} that the \DaD configuration tends to stretch to a smaller radius rather then be pulled to infinite $r$. Although absent in the $r_0 \gg r_\epsilon$ regime, the $SU(2)$-charged  instability could theoretically develop when $r_0\sim r_\epsilon$, but without any evidence for such a behavior we consider this possibility to be unlikely.

In what follows we focus on the stability of the \DaD brane in the KW case. We start with the singular conifold geometry
\bea
\label{metric}
\d s^2=\d r^2+r^2\left({g_5^2\over 9}+{\d\theta^2+\sin^2\theta\, \d\varphi^2\over 9}+{\d\tilde\theta^2+\sin^2\tilde\theta \, \d\phi^2\over 6}\right),
\eea
where
\bea
g_5=\d\psi+\cos\theta\, \d\varphi+\cos\tilde\theta \,\d \phi\, .
\eea
and consider a static configuration
\bea
\tilde\theta & = &\tilde\theta(\phi,\psi,\theta,\varphi)\,, \qquad r =  r(\phi,\psi,\theta,\varphi)\, .
\eea
The classical configuration found in~\cite{KSII} corresponds to
\bea
\label{clsol}
\tilde\theta={\pi\over 2}\, ,  \qquad
\cos\left({4\over \sqrt{6}}\,{\phi}\right)={r_0^4\over r^4}\, .
\eea
We want to show that this configuration minimizes~(\ref{stabfunct}) at least locally.

A comment is in order. Instead of using the radial coordinate $r$ as an independent variable, we are using ${\phi}$. This is because of the singularity of the $r$ coordinate at the tip of the \DaD profile. Use of $\phi$ helps to avoid this problem, since the point $\phi=0$ is regular. The price we pay is mild: the boundary conditions for the $SU(2)$-invariant mode of $r$ are such that it diverges at the end points ${\phi}_0=\pm {\sqrt{6}\pi/ 8}$ as $r\sim ({\phi}-{\phi}_0)^{-1/4}$, but this will not cause us much trouble. The $SU(2)$-invariant modes were already analyzed in~\cite{KSII} using a different coordinate system. Our goal is to focus on the $SU(2)$-charged modes of $r$ and $\tilde\theta$, which vanish at the endpoints $\phi=\pm {\sqrt{6}\pi/ 8}$.

For obvious convenience reasons we introduce $\alpha\in [-{\pi/2},{\pi/2}]$ through ${\phi}={\sqrt{6}}\alpha/4$
and expand the profiles $\tilde\theta$ and $r$ as follows
\bea
\tilde\theta &=& {\pi\over 2}+ \Theta(\alpha,\psi,\theta,\varphi)\, ,\\
r &=& {r_0\over\cos^{1/4}\alpha}+R(\alpha,\psi,\theta,\varphi)\, .
\eea
The boundary conditions for $\Theta$ and $R$ are such that they vanish at the endpoints $\alpha=\pm{\pi/2}$ except for the mode of $R$, which is independent of $\psi$, $\theta$ and $\varphi$. Plugging this into~(\ref{metric}) and expanding~(\ref{stabfunct}) up to the quadratic order we get
\begin{multline}
\label{seff}
S_{\rm eff}=\int \d\alpha\, \d\psi\, \d\theta\, \d\varphi \left({r_0^2\sin\theta \cos^{1/2}\alpha\over 144}\left(16{R'}^2\cos^2\alpha+9R^2\cos^2\alpha-5R^2\right)\right. +
\\ \left. +{r_0^4\sin\theta\over  144}\left({8\over 3}{\Theta}'^2-\Theta^2\right)+{2 r_0^3\sqrt{6}\sin\alpha\sin\theta \over 144 \cos^{3/4 }\alpha}{\partial R\over \partial \psi}\Theta\right. +
\\ + \left.
{r_0^2\cos^{1/2}\alpha\over 24\sin\theta}(\nabla R)^2+{r_0^4\over 144 \sin\theta\cos^2\alpha}(\nabla \Theta)^2
\right) .
\end{multline}
Here we introduced the notations
\bea
(\nabla f)^2={\partial f\over \partial \xi^i}{\partial f\over \partial \xi^j}g^{ij}\, ,\qquad \xi^i=\{\varphi,\theta,\psi\}\, ,\\
g^{ij}=\left(\begin{array}{ccc}
1 & 0 & -\cos\theta\\
0 & \sin^2\theta & 0 \\
-\cos\theta & 0 & 1+{1\over 2}\sin^2\theta
\end{array}\right).
\eea

Let us first say few more things about~(\ref{seff}) in the case when the perturbations $R$ and $\Theta$ have no dependence on the angular variables $\psi$, $\theta$ and $\phi$. The variation of the classical solution~(\ref{clsol}) with respect to $r_0$ gives the zero mode $R=(\cos\alpha)^{-1/4}$ of the corresponding equation
\bea
\label{Req}
-16{\d\over \d\alpha}\left(\cos^{5/2}\alpha{\d R\over \d\alpha}\right)+9\cos^{5/2}\alpha\, R-5\cos^{1/2}\alpha \, R=\lambda\cos^{3/2}\alpha\,R\, .
\eea
Clearly, presence of the zero mode implies that the first line of~(\ref{seff}) is not positively defined. Obviously the problematic term is $-5\cos^{1/2}\alpha R^2$. Later we will show that the angular-dependent term cures this problem. In the angular-independent case the absence of the negative modes can be derived from the relation between the number of zeros
and the principal quantum number of the wavefunction of the one-dimensional quantum-mechanical problem (\ref{Req}). We would also like to note here that the equation~(\ref{Req}) is a Heun equation, which admits an extensive symmetry group.

The effective action for the $SU(2)$-independent modes of $\Theta$,
\bea
\label{thetaaction}
{8\over 3}\,{\Theta'}^2-\Theta^2\, ,
\eea
is obviously positively defined on the space of functions vanishing at $\alpha=\pm {\pi/2}$. There are also zero modes
\bea
\label{zmode}
\Theta=\sin(\phi+\Delta\phi)=\sin\left({4\over \sqrt{6}}\,\alpha+\Delta\phi\right),
\eea
which break boundary conditions. These zero modes correspond to the rotation of $S^2$ by the broken $SU(2)$ symmetry of the conifold. The constant $\Delta\phi$ depends on a choice of the axis of the $U(1)\subset SU(2)$ rotation of $S^2$.

Now we introduce the angular dependence and would like to show that~(\ref{seff}) is positively defined.
First we would like to estimate from below the contribution of
\bea
\label{ang}
{r_0^2\over 144}\int \d\varphi\, \d\psi\, {\d\theta\over \sin\theta}\
\left(6\cos^{1/2}\alpha(\nabla R)^2+{r_0^2\over \cos^2\alpha}\,(\nabla \Theta)^2 \right).
\eea
First we expand $R$ and $\Theta$ in angular harmonics (recall that $\varphi$ has periodicity $2\pi$ and $\psi$ has periodicity $4\pi$):
\bea
R={1\over \sqrt{2}}\sum_{n_\varphi,n_\psi}R_{n_\varphi,n_\psi}(\alpha,\theta)\, \text{e}^{i(n_\psi \psi+n_\varphi \varphi)}\ ,\qquad
R_{n_\varphi,n_\psi}=R_{-n_\varphi,-n_\psi}^*\, ,
\eea
and similarly for $\Theta$. Here $n_\varphi$ and $n_\psi$ are either both integer or half-integer. To estimate (\ref{ang}) we need to find the ground state of the corresponding eigenvalue problem
\bea
\label{eigenpr}
-{1\over \sin\theta}{\d\over \d\theta}\left(\sin\theta\, {\d\Psi\over \d\theta}\right)+{(n_\varphi^2+n_\psi^2-2n_\varphi n_\psi\cos\theta)\over \sin^2\theta}\,\Psi=\lambda\,\Psi\, .
\eea
Assuming $n_\varphi,n_\psi>0$ the  corresponding ground state is
\bea
\Psi=(\cos\theta)^{n_\varphi+n_\psi}(\sin\theta)^{|n_\varphi-n_\psi|} \,,
\eea
and the corresponding eigenvalue is $\lambda=n^2+n$, where $n={\rm max}\{n_\varphi,n_\psi\}$. Plugging this back we obtain
\begin{multline}
\label{effactionII}
{16\pi^2 r_0^2\over 144} \int^{\pi\over 2}_{-{\pi\over 2}} \d\alpha\
{\cos^{1/2}\alpha}\left(16 R_{1/2,1/2}'^2\cos^2\alpha+9R_{n_\varphi,n_\psi}^2\cos^2\alpha-5R_{n_\varphi,n_\psi}^2\right)+
\\
+{r_0^2 }\left({8\over 3}\,\Theta_{n_\varphi,n_\psi}'^2-\Theta_{n_\varphi,n_\psi}^2\right) +
\\ +
  \left(6\cos^{1/2}\alpha|R_{n_\varphi,n_\psi}|^2+{r_0^2\over \cos\alpha^2}|\Theta_{n_\varphi,n_\psi}|^2\right) \left(
n^2+2n -n_\psi t\cos\gamma \sin\alpha \right),
\end{multline}
where
$$
t={2\sqrt{6}r_0|R_{n_\varphi,n_\psi}\Theta^*_{n_\varphi,n_\psi}|\over \cos^{3/4}\alpha\left(6\cos^{1/2}\alpha|R_{n_\varphi,n_\psi}|^2+{r_0^2\over \cos\alpha^2}|\Theta_{n_\varphi,n_\psi}|^2\right)}\ ,~ \cos\gamma ={{\rm Im}(R_{n_\varphi,n_\psi}\Theta^*_{n_\varphi,n_\psi})\over |R_{n_\varphi,n_\psi}\Theta^*_{n_\varphi,n_\psi}|}\,.
$$
Obviously $t$, which can always be chosen to be positive, is never larger than $1$. The term with $t$ in~(\ref{effactionII}) is negative. It causes the largest possible negative contribution when $\cos\gamma=1$ and $n=n_\psi$, i.e. $n_\psi \geq n_\varphi$. Since $t\sin\alpha<1$ the effective action~(\ref{effactionII}) is positive whenever $-5+6(n^2+n)>0$. Hence there is only one tricky case to analyze: $n=n_\varphi=n_\psi=1/2$ and $\cos\gamma=1$.

Let us first analyze this case assuming that the perturbation of $\theta$ is zero, that is $t=0$. Then we have $-5+6\cdot({1\over 2^2}+2{1\over 2})>0$ and the effective action is positive. What happens when $t\neq 0$? In this case, naively, the action for $R$ is not positively defined. If $R'$ is sufficiently small near the endpoints $\alpha=\pm{\pi/2}$, one can expect a negative contribution. However there is a positive contribution coming from $\Theta$-dependent terms that overpower the negative contribution from $R$. To make the analysis more quantitative we estimate~(\ref{effactionII}) from below as follows. First we drop the non-negative term $R'^2$ and substitute for~(\ref{thetaaction}) its minimal value $5\Theta^2/3$. Then we express $\Theta$ in terms of $R$ and $t$. Eventually we obtain
\begin{multline}
\label{effactionIII}
{16\pi^2 r_0^2\over 144}\sum_{n_\phi,n_\psi} \int^{\pi\over 2}_{-{\pi\over 2}} \d\alpha\ R_{n_\varphi,n_\psi}^2 \left(
{\cos^{1/2}\alpha}\left(9\cos^2\alpha-5\right)+10\cos^{5/2}\alpha \left({1-\sqrt{1-t^2}\over t}\right)^2 \right. \\  \left. +
12\cos^{1/2}\alpha\left({1-\sqrt{1-t^2}\over t^2}\right) \left({5\over 4}- {1\over 2}\, t \sin\alpha \right)\right).
\end{multline}
This is positive for all $0\leq t\leq 1$ and $-{\pi/ 2}\leq \alpha \leq{\pi/ 2}$, which completes the proof.


\section{Size of the baryon}
\label{sizeofbaryon}

In the leading approximation we can calculate the mass of the baryon by evaluating the tension of the D3-brane wrapping the minimal $S^3$ near the tip of the conifold. Indeed, as we argue in section~\ref{svi}, the D3 will prefer to sit at the IR end of the geometry for all values of $r_0$. Thus the leading mass of the baryon is always $M\Lambda$, cf.~(\ref{massofbaryon0}).

Let us discuss corrections to the mass in the regime, when the D3 dissolves in the \DaD to become a flavor instanton. To find corrections and estimate the size of the baryon we should find a static instanton configuration on the \DaD. The procedure is very similar to the calculation done by Sakai and Sugimoto in the case of the \DaDe branes in the type IIA theory~\cite{SSb}. The effective theory for the $U(N_f)$ gauge fields on the \DaD is given by the DBI part~(\ref{DBI eff}) and the Chern-Simons part~(\ref{CScoupling}). We split the gauge fields in the $U(1)$ and $SU(N_f)$ parts~(\ref{UNtoSUN}). The non-abelian gauge field $A$ will form an instanton configuration charged under the ``electric'' field $\hat{A}_0$. At first we ignore the Chern-Simons term, although it is non-trivial due to $M$ units of the RR flux through $S^3$, and set the $U(1)$ gauge field to zero $\hat{A}=0$.

The action for the non-abelian fields $A$ follows from the expression~(\ref{DBI eff}):
\bea
\label{action0}
\frac{1}{4g_s \alpha'^2}\int \d^4 x\, \d z\ k(z) \Tr\left[ h F_{\mu\nu}^2+2 F_{\mu z}^2\right],
\eea
where $k(z)$ is given by~(\ref{k(z)}) and the a radial coordinate $z$ was introduced according to~(\ref{zmetric}). In terms of this coordinate the geometric profile is $z\rightarrow -z$ symmetric and has a parabolic behavior $r(z)=r_0+O(z^2)$ near the tip.

We are interested in the static ($x_0$-independent) instanton solution with $F_{0i}=F_{0z}=0$. Employing the usual trick we find that the mass of such a state is not smaller than
\bea
\label{action1}
m_{\rm baryon}\geq \frac{1}{4g_s \alpha'^2}\int \d^3 x\, \d z\ k(z) h^{1/2}(z) \left|\epsilon^{ijk}\Tr F_{ij}F_{kz}\right|\, .
\eea

Since $k(z)h^{1/2}(z)\sim r$ is a monotonously increasing function of $z$ we conclude that the energy~(\ref{action0}) is minimal for an infinitely small instanton localized at $z=0$. The mass of such a configuration exactly matches the mass of the D3-brane wrapping the minimal $S^3$~(\ref{massofbaryon0}). For an instanton of the small size $\rho$, such that $\rho \ll r_\epsilon$ ($r_\eps\equiv \eps^{2/3}$) the leading correction to the mass will be of the form
\bea
\label{corr1}
\delta m_{\rm baryon}^{(1)}= m_{\rm baryon}^{(0)}\,{a \rho^2\over \eps^{4/3}} \, ,
\eea
where  $a$ is some positive coefficient of order $1$.

Similarly to the logic of \cite{SSb} so far we ignored the $U(1)$ gauge field and possible contributions from the Chern-Simons term~(\ref{CScoupling}):
\bea
\label{CS}
\frac{M}{24\pi^2}\int \omega_5({\cal{A}}) = - \frac{M}{16\pi^2}\int \d^4x\, \d z\  \hat{A}_0 {\epsilon^{ijk}\over 2}\,\Tr F_{ij}F_{kz} +...
\eea
The instanton carries the $U(1)$ charge and sources the electrostatic field. To calculate the corresponding contribution to the baryon mass we will follow the same procedure as in the \DaDe case. We assume that the size of the instanton $\rho$ is very small compared with $r_\eps$ and therefore we can expand the action around $z=0$ effectively reducing the geometry to the flat space. It is also convenient to rescale the spatial coordinates $x_i\rightarrow y_i=h^{-1/2}(r_\eps) x_i$ bringing~(\ref{action0}) to the canonical form such that $\rho$ now is also the spatial size of the instanton in coordinates $y_i$. We ignore the precise values of the numerical coefficients in what follows. Focusing on the static configuration with only $\hat{A}_0\neq 0$, we have the following effective action
\bea
\label{electric}
M \int \d x^0 \int
\d^3 y\,\d z  \left(C_1\Lambda\, h(r_\eps)\left(\partial_i \hat{A}_0^2+\partial_z \hat{A}_0^2\right)+ C_2\hat{A}_0 \,\epsilon^{ijk}\, \Tr F_{ij}F_{kz}\right).
\eea

The energy of the charge of size $\rho$ scales like $1/\rho^2$ leading to yet another correction toward baryon's mass
\bea
\label{corr2}
\delta m_{\rm baryon}^{(2)}=m_{\rm baryon}^{(0)} \,{b \over h(r_\eps) (\Lambda \rho)^2}\, .
\eea
Here $b$ is some positive coefficient of order $1$. Combining this with the correction due to curvature (\ref{corr1}) we obtain for $\rho\ll r_\eps$
\bea
m_\text{baryon}=M \Lambda \left(1+a{\rho^2\over \epsilon^{4/3}}+b{\epsilon^{4/3} \over \lambda^2 \rho^2}\right)+...\ .
\eea
The energy reaches its minimum for
\bea
\label{size2}
\rho \sim {\epsilon^{2/3}\over \lambda^{1/2} }\, .
\eea
Let us analyze this result. First, the condition $\rho\ll r_\eps$ holds for large $\lambda$, hence our assumption that $\rho$ is small is self-consistent. Similarly to the Sakai-Sugimoto case the size of the baryon scales like $1/\sqrt{\lambda}$ and we cannot neglect higher $\alpha'$ corrections as their contribution is of order $h^{-1/2}\alpha' \rho^{-2}\sim 1$. That is why we are not interested in calculating the explicit values for the coefficients $a$ and $b$, since they will be corrected after the higher order terms are taken into account. Yet, we expect the instanton size to remain of the order given by equation~(\ref{size}). The instanton cannot shrink too much due to the large Coulomb energy. Similarly, it cannot significantly expand because in this case the energy will increase with the increase of the warped volume of $S^3$.


\end{appendix}


\end{document}